\documentclass[11pt,a4paper]{article}

\usepackage{amsmath,amssymb,graphicx,cite}

\newcommand{\mD}{\mathcal{D}}

\begin{document}

\title{Charged thin shells in unimodular gravity} 
\author{Gabriel R. Bengochea\thanks{e-mail: gabriel@iafe.uba.ar}, Ernesto F. Eiroa\thanks{e-mail: eiroa@iafe.uba.ar}, Griselda Figueroa-Aguirre\thanks{e-mail: gfigueroa@iafe.uba.ar}\\
{\small  Instituto de Astronom\'{\i}a y F\'{\i}sica del Espacio (IAFE, CONICET-UBA),}\\
{\small Ciudad Universitaria, 1428, Buenos Aires, Argentina}} 
\date{}
\maketitle

\begin{abstract}

In this article, we construct a broad family of spacetimes with spherically symmetric thin shells in unimodular gravity. We present the framework for the analysis of the dynamical stability of the configurations under perturbations preserving the symmetry. In particular, we consider two different scenarios in which the non-conservation of the energy-momentum tensor is allowed; in both the spacetime has a thin shell with charge surrounding vacuum. Our constructions avoid the existence of event horizons and singularities. In both cases, we have obtained stable configurations for suitable values of the parameters. We compare our results with those corresponding to general relativity, finding some differences in the matter content and in the stability regions in the parameter space.

\end{abstract}

\section{Introduction}\label{intro}

Unimodular gravity (UG) is an alternative theory of gravity closely related to general relativity (GR), which was first considered by Einstein in 1919 \cite{einstein1919}. In such an approach the gravitational field is described by the trace-free Einstein equations that can be derived from an action, where a fixed non-dynamical 4-volume element appears, see e.g. \cite{buch,unruh,henneaux,ellis2011,bengo23}. The presence of this 4-volume background can be interpreted as something that breaks the diffeomorphism invariance of GR, turning UG into a theory invariant only under volume-preserving diffeomorphisms. An interesting feature of UG, which became more popular after the work \cite{weinberg89}, is that within the framework of this theory of gravity, the cosmological constant simply plays the role of an integration constant linked to initial conditions, but it allows to decouple it from a possible vacuum quantum energy, since the latter does not gravitate in UG, see e.g. \cite{ellis2011}. In this way, the huge discrepancy of up to 120 orders of magnitude between the observed value of the cosmological constant and the value predicted by quantum field theory (QFT) \cite{carroll}, known as \emph{the cosmological constant problem}, finds an elegant solution that does not require new physics\footnote{Of course this does not solve the mystery of why it has the \emph{particular value} that it has today.}. But there is another feature of UG: the invariance of the matter action under the restricted volume-preserving diffeomorphisms brings the possibility of a non-conservation of the energy-momentum tensor, usually represented by a \emph{diffusion} term\footnote{Note that this same result can be obtained, without imposing such limitations, through a slightly different derivation. See Sect. 2.2 of \cite{bengo23}}. Within UG, the conservation becomes an extra hypothesis which, if assumed, recovers the GR field equations. Once the conservation of the energy-momentum tensor is adopted, UG is completely equivalent to GR and consistent with all observational tests. Only when one chooses not to impose this conservation is possible to find deviations from GR. For instance, in \cite{bonder23} it is shown that the non-conservation of the energy-momentum tensor implies the non-geodesic motion of pointlike particles, and in \cite{herrera24} it is concluded that UG has a well-posed initial value formulation.

In the last decade, interest in UG has resurged with special emphasis in the area of cosmology, see e.g. \cite{ellis14,nojiri16a,corral20,linares21,josset,perez19,perez21,fabris19,garcia19,barvi21,cesare22,fabris22a,amadei22,leon22,pia23}. In particular, misconceptions regarding the notion of diffeomorphism invariance (mainly its use in UG) and the choice of gauges in the treatment of cosmological perturbations, were considered and analyzed in depth in \cite{bengo23}. Compact objects \cite{astorga19} and spherically symmetric configurations \cite{fabris23} were recently explored in the context of UG; in the last case, the general equations for a static solution were introduced, and examples resulting from the non-conservation of the energy-momentum tensor were shown, in the presence of an electromagnetic field as well as of a self-interacting scalar field. Proposals discussing the construction of theories of gravity leading to a non-conservation of the energy–momentum tensor can be found for instance in \cite{velten21}. On the other hand, in \cite{maudlin20} the authors analyze the issue of energy conservation when the standard quantum setting and semiclassical gravity are addressed, concluding that probably neither Einstein equations nor conservation laws hold in nature. 
It is also interesting to mention that, within the context of QFT on curved spacetimes, considerations related to the issue of renormalization of the energy-momentum tensor in the semiclassical gravity framework \cite{wald95} also seem to suggest a preference for UG. Theoretical proposals coming from certain quantum gravity approaches suggest that, at the Planck scale, a more fundamental physics is expected to be discrete. For this approach to be compatible with Lorentz symmetry\footnote{Issues related to compatibility with Lorentz invariance when a discrete spacetime is considered, can be found in, e.g. \cite{collins04}.}, such a discreteness could be accomplished by some kind of 4-volume elementary building blocks.
These would produce a background 4-volume structure where only invariance under volume-preserving diffeomorphisms would be present, thus allowing UG to be considered as a natural effective description of gravity at low energies \cite{anderson71}. Also, Planck-scale discreteness could play a role in black hole physics \cite{perez17} and in a possible resolution of the information paradox associated with black holes \cite{perez15, amadei21}. For a recent discussion of these topics see e.g. \cite{perez22}. On the other hand, since the detection of the accelerated expansion of the universe, extensive studies are conducted to understand whether this acceleration is due to a cosmological constant, a dynamical dark energy, or a modification of the theory of gravity. The use of UG, with the perspective that the non-conservation of the energy-momentum is due to a fundamental granularity of the spacetime at Planckian scales, was implemented to search for answers about the nature of dark energy, the $H_0$ tension,  and the current value of the cosmological constant in recent works \cite{josset,perez19,perez21}, as well as to find a possible alternative for the inflationary phase (without an inflaton field) of the early universe \cite{amadei22, bengochea24}.

Thin shells of matter appear as idealized useful models in many physical contexts. In GR, the Darmois--Israel \cite{darmois1927,israel1966} junction conditions provide the tools for the construction of a new spacetime by joining two different geometries across a hypersurface. The formalism allows to analyze the characteristics and dynamics of thin shells, relating the energy--momentum tensor at the matching hypersurface to the geometries at both sides of it. The method has been broadly applied in many situations because of its simplicity and flexibility; the stability analysis is easy to perform in case of highly symmetric configurations, for perturbations preserving the symmetry. Many researchers have adopted the junction conditions to model vacuum bubbles and thin layers around black holes \cite{brady91,ishak02,goncalves02,lobo05,eiroa11,sharif15}, fluid spheres supported by thin shells \cite{rosa20}, wormholes \cite{poisson95,eiroa04a,eiroa08a,montelongo12,forgani18,berry20,
mazha14,eiroa19}, and gravastars \cite{gravstar04,gravstar06,gravstar12,das24}. There are also studies in which this formalism is used to build wormholes and thin shells of matter in $N$ dimensions \cite{rahaman06,dias10,eiroa12,banerjee18}. The junction conditions have been obtained in some theories of modified gravity. Different physical scenarios have been considered within $F(R)$ gravity in four dimensions by using this technique \cite{tswhFR16a,tswhFR20,tsFR17,tsFR18,tsFR19}, and also in lower \cite{tsFR3d21a,tsFR3d21b} and in higher \cite{tsFRNd22} dimensionality. Spacetimes with thin shells were also analyzed within Palatini $F(R)$ gravity \cite{lobo20}, in $F(R,T)$ gravity \cite{rosa21}, and in Brans-Dicke theory \cite{eiroa08b,eiroa10b}.

In this article, we study thin shells within UG with the help of the corresponding junction conditions introduced in \cite{ellis2011}, using spacetimes in which the non-conservation of the energy-momentum tensor is allowed. To our knowledge, this topic has not been previously explored in the literature. We start in Sec. \ref{ug-bh} with a review of the main aspects of the theory and the recently found spherically symmetric black hole solutions in the presence of the electromagnetic field \cite{fabris23}. In Sec. \ref{thin-shells}, we construct spherically symmetric thin shells within UG and we obtain the condition for stability under radial perturbations. In Sec. \ref{bubbles}, we provide examples of charged thin shells surrounding vacuum (bubbles), with flat, de Sitter, and anti-de Sitter asymptotics. Finally, in Sec. \ref{conclu}, we present the conclusions of this work. Throughout the paper, we use the $(-,+,+,+)$ signature for the spacetime metric and we adopt units such that $c=G=1$.

\section{Black holes in unimodular gravity}\label{ug-bh}

Let us start this section with the field equations of  UG, which can be obtained from a variational principle, requiring the extremization $\delta S=0$, with the action written as \cite{bengo23}\footnote{To avoid some of the misconceptions mentioned in \cite{bengo23}, we will first use indices following Wald's convention and notation for the geometrical objects \cite{wald84}, which makes a distinction between index notation and component notation. Then, we will denote components of a tensor in a given basis by using Greek indices.},
\begin{eqnarray}\label{accionUG}
	S[g^{ab}, \Psi_M;\Lambda] &=& \frac{1}{2 \kappa} \int \left[R \epsilon_{abcd}^{(g)} -2 \Lambda (\epsilon_{abcd}^{(g)} - \varepsilon_{abcd}) \right]+S_M[g^{ab}, \Psi_M],
\end{eqnarray}
where we define $\kappa \equiv 8 \pi$, $R$ is the Ricci scalar, and $\varepsilon_{abcd}$ and $\epsilon_{abcd}^{(g)}$ are a fiduciary 4-volume element (given by the theory) and a 4-volume element associated with the metric $g_{ab}$, respectively. The scalar $\Lambda(x)$ is a Lagrange multiplier function, and $S_M$ is the action of the matter fields generically represented by $\Psi_M$. Variations of \eqref{accionUG} with respect to dynamical variables $g^{ab}$, $\Lambda$, and $\Psi_M$ lead to
\begin{equation}\label{EFE}
	R_{ab} - \frac{R}{2}g_{ab} + \Lambda(x) g_{ab} = \kappa T_{ab} ,
\end{equation}
\begin{equation}\label{constraint0}
	\epsilon_{abcd}^{(g)}  = \varepsilon_{abcd} ,
\end{equation}
\begin{equation}\label{KG}
	\frac{\delta S^M}{\delta \Psi_M} = 0.
\end{equation}
On the right-side of Eqs. \eqref{EFE} the energy-momentum tensor appears, which is defined as
\begin{equation}\label{defTab}
	T_{ab} \equiv  \frac{-2}{\sqrt{-g}}  \frac{\delta S_M}{\delta g^{ab}} .
\end{equation}
Note that in the last expression $g$ is the determinant of the components of the metric tensor $g_{\mu \nu}$, in a specific coordinate basis. On the other hand, Eq. \eqref{KG} yields the equation of motion of the matter fields (i.e. a Klein-Gordon type equation).

At this point, the trace of Eq. \eqref{EFE} allows us to write $\Lambda$ as
\begin{equation}\label{lambda}
	\Lambda = \frac{\kappa T + R}{4},
\end{equation}
where $T = g^{ab} T_{ab}$ is the trace of the energy-momentum tensor. Next, by substituting the former expression into Eq. \eqref{EFE}, it leads to the trace-free part of Einstein field equations, namely
\begin{equation}\label{UGecs}
	R_{ab} - \frac{1}{4}  g_{ab} R = \kappa \left(T_{ab} - \frac{1}{4}  g_{ab} T \right) ,
\end{equation}
which are the UG equations for the gravitational field. Note that anything behaving like a cosmological constant or vacuum energy does not gravitate, since it automatically satisfies that the right-hand side of \eqref{UGecs} is zero.

As we mentioned in the Introduction, one of the interesting features of UG that allows departures from GR is the possibility of the non-conservation of the energy-momentum tensor; indeed, Eq. \eqref{EFE} lets $\nabla^a T_{ab}\ne 0$. This can be demonstrated rigorously in alternative ways, noting that the important point is that, as the theory is presented through action \eqref{accionUG}, UG has a non-dynamical element that can lead to the non-conservation of $T_{ab}$. In UG, $g^{ab}$, $\Lambda$ and $\Psi_M$ are dynamical variables, while the 4-volume element $\varepsilon_{abcd}$ is fixed and non-dynamical. One can choose to consider the variation of the action \eqref{accionUG} involving all the geometric objects, that is, applying diffeomorphisms on both dynamical and non-dynamical variables or, alternatively, although the action \eqref{accionUG} (by construction) is invariant under generic one-parameter family of diffeomorphism, one may restrict the consideration to the volume preserving diffeomorphisms when performing the variation of the matter action $S_M$. Both paths lead to \cite{bengo23}
\begin{equation}\label{conservTab}
	\nabla^a (T_{ab} - g_{ab} \mD) =0,
\end{equation}
where $\mD(x)$ is an arbitrary scalar field (i.e. the \emph{diffusion term}) that encapsulates the possibility of non-conservation.

From Eq. \eqref{conservTab}, applying the covariant derivative $\nabla^a$ to both sides of Eq. \eqref{EFE} and using the Bianchi's identities $\nabla^a [R_{ab}-\frac{1}{2}g_{ab} R]=0$, we obtain that
\begin{equation}\label{lambdax}
	\Lambda(x) = \Lambda_0 + \kappa \mD(x),
\end{equation}
where $\Lambda_0$ is simply a constant of integration, fixed by initial conditions. Note that, the case $\mD(x)=constant$ leads to the standard conservation law for $T_{ab}$. In the present work, we will be analyzing some particular forms for $\Lambda(x)$ motivated by the recent results of \cite{fabris23}.

In the article \cite{fabris23}, spherically symmetric solutions of the unimodular field equations were found, with a line element taking the form
\begin{equation}
ds^2 = -A(r)dt^2 + A^{-1}(r)dr^2+r^2(d\theta^2+\sin^2\theta d\varphi^2),
\label{metric_fabris}
\end{equation}
where the usual Schwarzschild $(t,r,\theta,\varphi)$ coordinates are used. Following \cite{fabris23}, we adopt two different functions $\Lambda(r)$, corresponding to two distinct behaviors both asymptotically and in the center ($r = 0$), and we include a radial electric field $E(r)$ associated with a charge $Q$.\footnote{There are typos in the solution presented in \cite{fabris23}, the correct expressions are shown here (private communication with J. C. Fabris).}

\subsection*{Case A}

The first choice is the power law 
\begin{equation}
\Lambda(r) = \Lambda_0 + \Lambda_1 r^p, 
\end{equation}
with $\Lambda _1$ a constant, for which the field equations have the solution \cite{fabris23} determined by
\begin{itemize}
\item $p\neq-4,-3$: 
\begin{equation}
A(r)= 1-\frac{2M}{r}+\frac{Q^2}{r^2}-\frac{\Lambda _0}{3}r^2-\frac{4 \Lambda _1}{(p+3)(p+4)}r^{p+2} ,
\label{metA1}
\end{equation}
\begin{equation}
E^2(r) = \frac{Q^2}{r^4} -\frac{p}{p+4}\Lambda_1 r^p ;
\label{efA1}
\end{equation}
\item $p=-4$: 
\begin{equation}
A(r)= 1-\frac{2M}{r}+\frac{Q^2}{r^2}-\frac{\Lambda _0 r^2}{3}+\frac{4 \Lambda _1}{r^2} \ln r ,
\label{metA2}
\end{equation}
\begin{equation}
E^2(r) = \frac{Q^2}{r^4} +\frac{4\Lambda_1}{r^4}\ln r ;
\label{efA2}
\end{equation}
\item $p=-3$: 
\begin{equation}
A(r)= 1-\frac{2M}{r}+\frac{Q^2}{r^2}-\frac{\Lambda _0 r^2}{3}-\frac{4 \Lambda _1}{r} \ln r ,
\label{metA3}
\end{equation}
with the squared electric field given by Eq. \eqref{efA1}. 
\end{itemize}
It is important to note that:
\begin{itemize}
\item $p=-4$: this case is clearly pathological since the electric field becomes imaginary near $r = 0$ when $\Lambda_1 > 0$, or for large $r$ if $\Lambda_1 < 0$. 
\item $p=0$: this case reduces to the GR solution (Reissner-Nordstr\"om with a cosmological constant) after the redefinition $\Lambda_0 \to \Lambda_0+\Lambda_1$.
\item $p=-2$: in this particular case, the metric function can be rearranged in the form
\begin{equation*}
A(r)= 1-2\Lambda _1-\frac{2M}{r}+\frac{Q^2}{r^2}-\frac{\Lambda _0 r^2}{3} ,
\end{equation*} 
so, after an appropriate change of coordinates, the spacetime presents an angular deficit if $\Lambda_1 > 0$ or an angular surplus when $\Lambda_1 < 0$.
\end{itemize}

\subsection*{Case B}

For the second choice
\begin{equation}
\Lambda(r) = \Lambda_0 + \frac{\Lambda_1 }{\left( r^2 +b^2 \right) ^2}, 
\end{equation} 
where $b$ is a constant, the field equations have the solution \cite{fabris23} given by
\begin{equation}
A(r)= 1-\frac{2M}{r}+\frac{Q^2}{r^2}-\frac{\Lambda _0}{3}r^2-\frac{2 \Lambda _1}{r^2}\left[ \frac{r}{b}\arctan \left( \frac{r}{b}\right)  - \ln \left( 1+\frac{r^2}{b^2} \right) \right] ,
\label{metB}
\end{equation}
\begin{equation}
E^2(r) = \frac{Q^2}{r^4} +\frac{\Lambda _1}{r^4} \left[\frac{b^2 \left(3 b^2+4 r^2\right)}{\left(b^2+r^2\right)^2}+2 \ln \left(1+\frac{r^2}{b^2}\right)\right]. 
\label{efB}
\end{equation}
Note that no change in the sign of $E^2(r)$ can be assured by taking $\Lambda _1 > 0$.

In both cases, for small enough values of charge, the spacetime corresponds to a black hole with a singularity at the center; as the charge grows beyond the extremal value, the event horizon vanishes and the singularity is naked \cite{fabris23}. In particular, when $\Lambda _1=0$ both solutions reduce to the Reissner-Nordstr\"om with a cosmological constant $\Lambda _0$ spacetime of GR.

\section{Spherical shells: construction and stability}\label{thin-shells}

In this section we study spherically symmetric thin shells, where a layer of matter appears as the result of cutting and pasting two manifolds at a surface in order to construct a new manifold. We start from the geometries
\begin{equation}
ds_{1,2}^2 = -A_{1,2}(r_{1,2})dt_{1,2}^2 +A_{1,2}^{-1}(r_{1,2})dr_{1,2}^2+r_{1,2}^2(d\theta^2+\sin^2\theta d\varphi^2),
\label{metric}
\end{equation}
and we take the spherical surface $\Sigma $ such as $r_{1,2}=a$. We define $\mathcal{M}_1$ as the set of points with radial coordinate $r_1\leq a$ and $\mathcal{M}_2$ the one with $r_2\geq a$. We join them at $\Sigma $, so the resulting manifold ${\mathcal M}$ is the union of the inner part $\mathcal{M}_1$ and the outer part $\mathcal{M}_2$. With a suitable identification, we can introduce a global radial coordinate $r$ in ${\mathcal M}$, with the surface $\Sigma $ located at $r=a$. For the study of the stability of our construction, we let the radius $a$ to be a function of the proper time $\tau$ measured by an observer at the joining surface. The line element is continuous across $\Sigma $ as the time coordinates in each side are chosen to satisfy $d\tau^2 = A_1(a)^2\left( A_1(a) + \dot{a}^2\right) ^{-1}dt_1^2=A_2(a)^2\left( A_2(a) + \dot{a}^2\right) ^{-1}dt_2^2$, where the dot stands for $d/d\tau $. The induced metric on $\Sigma$ then reads
\begin{equation}
ds_\Sigma^2=-d\tau^2+a^2(\tau)(d\theta ^2+\sin^2\theta d\varphi ^2).
\end{equation}
We denote the coordinates of the embedding by $X_{1,2}^\mu=(t_{1,2},r,\theta,\varphi)$ and the coordinates at the surface $\Sigma$ by $\xi ^i=(\tau , \theta,\varphi )$. The relation between the geometry and the matter on this surface in UG takes the form \cite{ellis2011} 
\begin{equation}
-\left[ K_{\mu\nu}-K\left( h_{\mu\nu}-\frac{1}{2}g_{\mu\nu}\right)\right] = 8\pi \left( S_{\mu\nu}-\frac{1}{4}S g_{\mu\nu} \right),
\label{junctcond}
\end{equation}
where $h_{\mu\nu}$ is the induced metric on $\Sigma$, $K_{\mu\nu}$ is the extrinsic curvature, $K$ is its trace, and $S_{\mu\nu}$ is the surface energy-momentum tensor; the brackets  $[\Upsilon]= \Upsilon^2\vert_\Sigma -\Upsilon^1\vert_\Sigma$ denote the jump of $\Upsilon$ across the surface. The geometry is continuous across $\Sigma$, i.e.  $[h_{\mu\nu}]=0$, as demanded by the junction formalism. If $[K_{\mu \nu }]=0$ we speak of $\Sigma$ as a boundary surface and if $[K_{\mu \nu }]\neq 0$ we say that there is a thin shell of matter at $\Sigma$. The general form of the components of  $K_{ij}$ at each side of $\Sigma $ are given by
\begin{equation}
K_{ij}^{1,2} = - n_{\gamma}^{1,2} \left. \left( \frac{\partial^2
X^{\gamma}_{1,2}}{\partial \xi^i \partial \xi^j} +\Gamma_{\alpha\beta}^{\gamma}
\frac{\partial X^{\alpha}_{1,2}}{\partial \xi^i} \frac{\partial
X^{\beta}_{1,2}}{\partial \xi^j} \right) \right|_{\Sigma}, 
\label{sff}
\end{equation}
where $n_{\gamma}^{1,2}$ are unit normals ($n^{\gamma} n_{\gamma} = 1$) to $\Sigma $. With the definition ${\mathcal H}(r)=r-a(\tau)=0$, they take the form
\begin{equation}
n_{\gamma}^{1,2} = \left| g^{\alpha \beta} \frac{\partial
\mathcal{H}}{\partial X^{\alpha}_{1,2}} \frac{\partial \mathcal{H}}{\partial
X^{\beta}_{1,2}} \right|^{- 1 / 2} \frac{\partial \mathcal{H}}{\partial
X^{\gamma}_{1,2}}, 
\label{normal1}
\end{equation}
where the unit normals at both sides of $\Sigma $ are oriented outwards from the origin. In this way, the normal to $\Sigma$ is unique and points from region 1 to region 2 as required by the sign convention used in Eq. (\ref{junctcond}). We adopt the orthonormal basis $\{ e_{\hat{\tau}}=e_{\tau }, e_{\hat{\theta}}=a^{-1}e_{\theta }, e_{\hat{\varphi}}=(a\sin \theta )^{-1} e_{\varphi }\} $ at the shell. 
Within this frame, for the metric (\ref{metric}), the first fundamental form reads $h^{1,2}_{\hat{\imath}\hat{\jmath}}= \mathrm{diag}(-1,1,1)$, the unit normal is
\begin{equation} 
n_{\gamma }^{1,2}= \left(-\dot{a},\frac{\sqrt{A_{1,2}(a)+\dot{a}^2}}{A_{1,2}(a)},0,0 \right),
\label{normal2}
\end{equation}
and the second fundamental form has the only non-null components
\begin{equation} 
K_{\hat{\theta}\hat{\theta}}^{1,2}=K_{\hat{\varphi}\hat{\varphi}}^{1,2
}=\frac{1}{a}\sqrt{A_{1,2} (a) +\dot{a}^2}
\label{sff1}
\end{equation}
and
\begin{equation} 
K_{\hat{\tau}\hat{\tau}}^{1,2 }=-\frac{A '_{1,2}(a)+2\ddot{a}}{2\sqrt{A_{1,2}(a)+\dot{a}^2}},
\label{sff2}
\end{equation}
where the prime stands for $d/dr$. We consider a conservative perfect fluid for the surface energy-momentum tensor, which in the orthonormal basis has the form $S_{_{\hat{\imath}\hat{\jmath} }}={\rm diag}(\sigma ,p,p)$, with $\sigma$ the surface energy density and $p$ the isotropic transverse pressure. From Eq. (\ref{junctcond}), with the help of Eqs. (\ref{sff1}) and (\ref{sff2}), we obtain
\begin{equation}
\sigma = -\frac{1}{4\pi a}\left( \sqrt{A_2(a)+{\dot a}^2}-\sqrt{A_1(a)+{\dot a}^2}\right) 
\label{sigma}
\end{equation}
and
\begin{equation}
p=-\frac{\sigma}{2}+\frac{1}{16\pi}\left( \frac{2\ddot a+A'_2(a)}{\sqrt{A_2(a)+{\dot a}^2}}-\frac{2\ddot a+A'_1(a)}{\sqrt{A_1(a)+{\dot a}^2}}\right) .
\label{pres}
\end{equation}
The expressions of $\sigma $ and $p$ have the same form as in GR, the reason seems to be that we have adopted a conservative surface energy-momentum tensor for the perfect fluid at the shell.
These two equations above, or any of them plus the equation
\begin{equation}
\frac{d(a^2\sigma)}{d\tau}+p\frac{da^2}{d\tau}=0,
\label{conserv}
\end{equation}
determine the evolution of the shell radius as a function of $\tau$. Considering that $\mathcal{A}=4\pi a^2$ is the area of $\Sigma $, the first term can be understood as the change of the internal energy $\epsilon = \sigma \mathcal{A}$, while the second one represents the work done by the pressure, so this equation provides an energy balance on the shell. With the help of a given equation of state $p=p(\sigma)$, we can formally integrate Eq. \eqref{conserv} to give $\sigma=\sigma(a)$. For the analysis of the mechanical stability of a static thin shell with radius $a_0$, we now consider small perturbations preserving the symmetry. In this static case, the energy density and the pressure are given by
\begin{equation}
\sigma _0 = -\frac{1}{4\pi a_0}\left( \sqrt{A_2(a_0)}-\sqrt{A_1(a_0)}\right) 
\label{sigma0}
\end{equation}
and
\begin{equation}
p_0=-\frac{\sigma _0}{2}+\frac{1}{16\pi}\left( \frac{A'_2(a_0)}{\sqrt{A_2(a_0)}}-\frac{A'_1(a_0)}{\sqrt{A_1(a_0)}}\right) .
\label{pres0}
\end{equation}
After some algebraic manipulations, from Eq. (\ref{sigma}) we obtain
\begin{equation}
\dot{a}^2+V(a)=0,
\label{pot1}
\end{equation}
where
\begin{equation}
V(a)=\frac{A_1(a)+A_2(a)}{2}-\left( 2\pi a \sigma (a) \right) ^2-\left( \frac{A_1(a)-A_2(a)}{8\pi a \sigma (a)}\right) ^2
\label{pot2}
\end{equation}
is commonly interpreted as a potential, given the analogy between Eq. (\ref{pot1}) and the energy of a point particle with only one degree of freedom. This potential can be expanded around the static solution, to give
\begin{equation}
V(a)=V(a_0)+V'(a_0)(a-a_0)+\frac{V''(a_0)}{2}(a-a_0)^2+\mathcal{O}(a-a_0)^3\ .
\end{equation}
It is straightforward to see that $V(a_0)=0$. The first derivative of the potential takes the form
\begin{eqnarray}
V'(a) &=&  \frac{A_{1}'(a)+A_{2}'(a)}{2}
-\frac{(A_{1}(a)-A_{2}(a)) \left(A_{1}'(a)-A_{2}'(a)\right)}{32 \pi ^2 a^2 \sigma (a)^2} \nonumber \\ 
&& +   \left( \sigma (a) + a \sigma '(a) \right) 
\left( \frac{(A_{1}(a)-A_{2}(a))^2}{32 \pi ^2 a^3 \sigma (a)^3} -8 \pi ^2 a \sigma (a) \right);
\end{eqnarray}
from Eq. (\ref{conserv}) we find that $a\sigma '(a) = -2(\sigma(a)+p(a))$, then
\begin{eqnarray}
V'(a) &=&  \frac{A_{1}'(a)+A_{2}'(a)}{2}
-\frac{(A_{1}(a)-A_{2}(a)) \left(A_{1}'(a)-A_{2}'(a)\right)}{32 \pi ^2 a^2 \sigma (a)^2} \nonumber \\ 
&& - \left( \sigma (a) + 2 p(a) \right)
\left( \frac{(A_{1}(a)-A_{2}(a))^2}{32 \pi ^2 a^3 \sigma (a)^3} -8 \pi ^2 a \sigma (a) \right) .
\end{eqnarray}
After some algebra we can verify that $V'(a_0)=0$. The second derivative of the potential reads
\begin{eqnarray}
\nonumber V''(a)&=& \frac{A_1''(a)+A_2''(a)}{2}-\frac{(A_1(a)-A_2(a)) \left(A_1''(a)-A_2''(a)\right)}{32 \pi ^2 a^2 \sigma (a)^2} \\ 
\nonumber && -\frac{\left(A_1'(a)-A_2'(a)\right)^2}{32 \pi ^2 a^2 \sigma (a)^2} +\left(a \sigma '(a)-2 p(a)\right)  \frac{(A_1(a)-A_2(a)) \left(A_1'(a)-A_2'(a)\right)}{16 \pi ^2 a^3 \sigma (a)^3} \\ 
\nonumber && + \left(\sigma (a) \left(-2 a p'(a)+6 p(a)+3 \sigma (a)\right)+2 a (3 p(a)+\sigma (a)) \sigma '(a)\right) \frac{(A_1(a)-A_2(a))^2 }{32 \pi ^2 a^4 \sigma (a)^4} \\ 
&& +8 \pi ^2 \left(\sigma (a) \left(2 a p'(a)+2 p(a)+\sigma (a)\right)+2 a (p(a)+\sigma (a)) \sigma '(a)\right).
\end{eqnarray}
We adopt at the shell the linearized equation of state 
\begin{equation}
p-p_0 =\eta (\sigma -\sigma_0)+\mathcal{O} (\sigma -\sigma_0)^2
\label{eos}
\end{equation}
where $\eta$ is a parameter that can be interpreted as the fluid squared velocity of sound if $0\le \eta <1$. Using again that $a\sigma '(a) = -2(\sigma(a)+p(a))$, the second derivative of the potential evaluated at $a_0$ reads
\begin{eqnarray}
\nonumber V''(a_0)&=& \frac{A_{1}''(a_0)+A_{2}''(a_0)}{2} -\frac{(A_{2}(a_0)-A_{1}(a_0)) \left(A_{2}''(a_0)-A_{1}''(a_0)\right)}{32 \pi ^2 a_0^2 \sigma _0^2} \\
\nonumber && -\frac{\left(A_{1}'(a_0)-A_{2}'(a_0)\right)^2}{32 \pi ^2 a_0^2 \sigma _0^2} - \left( \sigma _0 +2 p_0 \right) \frac{(A_{1}(a_0)-A_{2}(a_0)) \left(A_{1}'(a_0)-A_{2}'(a_0)\right)}{8 \pi ^2 a_0^3 \sigma _0^3}  \\ 
\nonumber && - \left(2 (5-2 \eta) p_0 \sigma _0+12 p_0^2+(1-4 \eta) \sigma _0^2\right) \frac{(A_1(a_0)-A_2(a_0))^2}{32 \pi ^2 a_0^4 \sigma _0^4} \\
&& -8 \pi ^2 \left(2 (2 \eta +3) p_0 \sigma _0 +4 p_0^2+(4 \eta +3) \sigma _0^2\right) ,
\end{eqnarray}
where $\sigma _0$ and $p_0$ are given by Eqs. (\ref{sigma0}) and (\ref{pres0}), respectively. 
By replacing the expressions of $\sigma _0$ and $p_0$, we can rewrite the second derivative of the potential in the form
\begin{eqnarray}
V''(a_0)&=& - \frac{\sqrt{A_2(a_0)} A_1''(a_0)-\sqrt{A_1(a_0)} A_2''(a_0)}{ \sqrt{A_1(a_0)}-\sqrt{A_2(a_0)}}+\frac{A_2(a_0)^{3/2} A_1'(a_0)^2-A_1(a_0)^{3/2} A_2'(a_0)^2}{2 A_1(a_0) A_2(a_0) \left(\sqrt{A_1(a_0)}-\sqrt{A_2(a)}\right)} \nonumber \\ 
&& -(2 \eta +1)\frac{ \sqrt{A_2(a_0)}\left( a_0 A_1'(a_0)- 2 A_1(a_0)\right) -\sqrt{A_1(a_0)}\left( a_0 A_2'(a_0)-2 A_2(a_0)\right) }{ a_0^2  \left( \sqrt{A_1(a_0)}-\sqrt{A_2(a_0)}\right)} .
\label{potD2VSinSigmaP}
\end{eqnarray}
The configuration is stable under radial perturbations when $V''(a_0)>0$.

\section{Charged shells surrounding vacuum}\label{bubbles}

\begin{figure}[t!]
\centering
\includegraphics[width=0.98\textwidth]{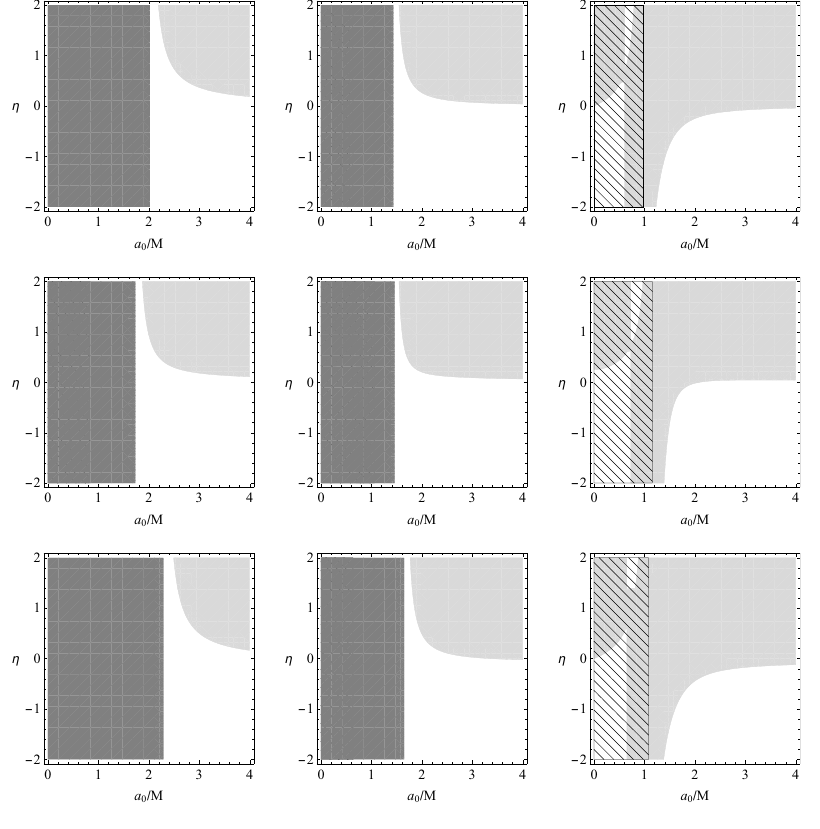}
\caption{Regions of stability in the $(a_0/M,\eta)$ plane for $\Lambda_0 M^2=0$. The top row displays the results corresponding to GR ($\Lambda_1 =0$), for which $Q_{c}/M=1$; the center row to UG case A with $p=-5$ and $\Lambda_1 M^{p+2}=-0.4$, for which $Q_{c}/M=0.55$; and the bottom row to UG case B with $b/M=1$ and $\Lambda_1 /M^2=0.4$, for which $Q_{c}/M=1.04$. The left column shows the plots with $Q=0$, the center column  with $|Q|=0.9Q_{c}$, and the right column with $|Q|=1.1Q_{c}$. Configurations in the light gray zone are stable, the dashed area represents thin shells that do not satisfy WEC, and the dark gray region is non-physical.}
\label{fig:fig1}
\end{figure}
\begin{figure}[t!]
\centering
\includegraphics[width=0.98\textwidth]{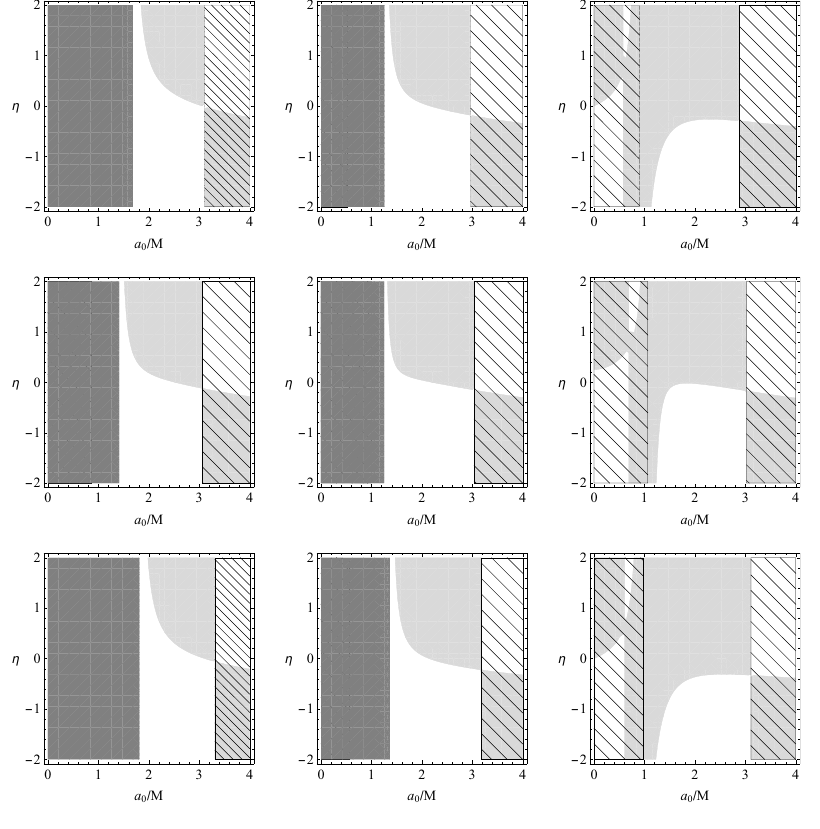}
\caption{Regions of stability in the $(a_0/M,\eta)$ plane for $\Lambda_0 M^2=-0.2$. The top row displays the results corresponding to GR ($\Lambda_1=0$), for which $Q_{c}/M=0.97$; the center row to UG case A with $p=-5$ and $\Lambda_1 M^{p+2}=-0.4$, for which $Q_{c}/M=0.41$; and the bottom row to UG case B with $b/M=1$ and $\Lambda_1 /M^2=0.4$, for which $Q_{c}/M=1.00$. The left column shows the plots with $Q=0$, the center column with $|Q|=0.9Q_{c}$, and the right column with $|Q|=1.1Q_{c}$. The meanings of the light gray, dark gray, and dashed zones are the same as in Fig. \ref{fig:fig1}.}
\label{fig:fig2}
\end{figure}
\begin{figure}[t!]
\centering
\includegraphics[width=0.98\textwidth]{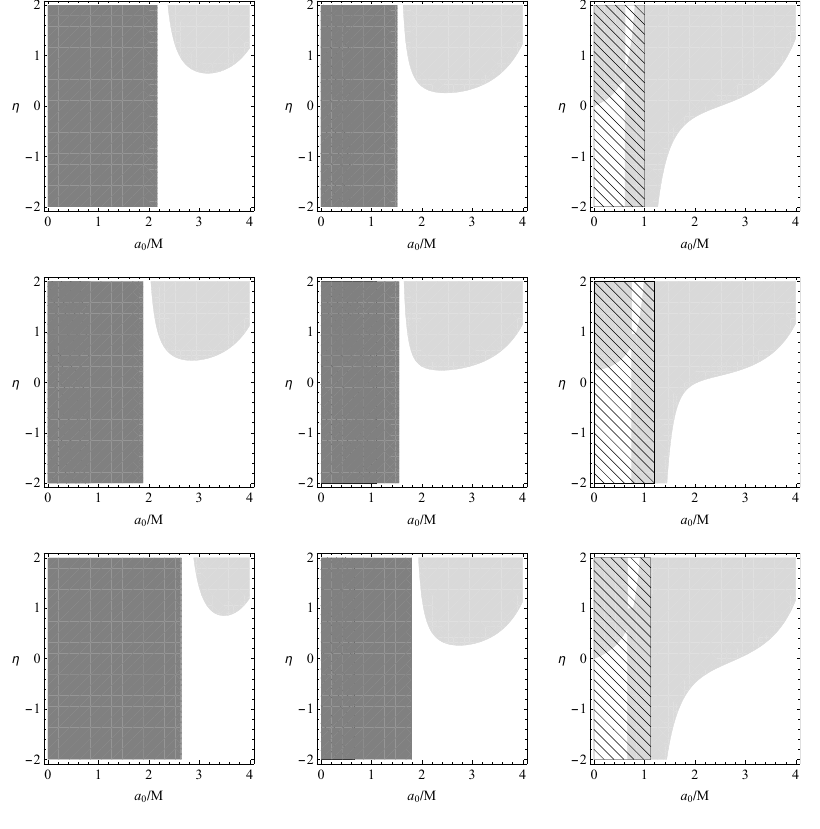}
\caption{Regions of stability in the $(a_0/M,\eta)$ plane for $\Lambda_0 M^2=0.05$. The top row displays the results corresponding to GR ($\Lambda_1 =0$), for which $Q_{c}/M=1.01$; the center row to UG case A with $p=-5$ and $\Lambda_1 M^{p+2}=-0.4$, for which $Q_{c}/M=0.59$; and the bottom row to UG case B with $b/M=1$ and $\Lambda_1 /M^2=0.4$, for which $Q_{c}/M=1.06$. The left column shows the plots with $Q=0$, the center column with $|Q|=0.9Q_{c}$, and the right column with $|Q|=1.1Q_{c}$.  The cosmological horizons are located at the radius (from left to right)  $r_c/M=6.43, 6.54, 6.60$ (top row), $r_c/M=6.45, 6.49, 6.50$ (center row), and $r_c/M=5.68, 5.89, 5.98$ (bottom row). The meanings of the light gray, dark gray, and dashed zones are the same as in Fig. \ref{fig:fig1}.}
\label{fig:fig3}
\end{figure}

Let us introduce two concrete examples. For the inner zone we adopt a Minkowski geometry, that is $A _1(r) = 1$, and for the outer region the UG spacetime with a radial electric field introduced in Sec. \ref {ug-bh}, determined by the metric function $A_2(r)$, given by Eqs. \eqref{metA1}, \eqref{metA2}, or \eqref{metA3} in the case A  and by Eq. \eqref{metB} in the case B. The possible horizons are the real and positive zeros of the function $A_2(r)$. In our construction, the radius $a_0$ of the thin shell is taken larger than the radius $r_h$ of the event horizon but smaller than the radius $r_c$ of cosmological horizon of the original manifold corresponding to the outer part, if any of them exist. In this way, we obtain a vacuum region surrounded by a charged thin shell, i.e. a charged bubble, without singularities and event horizons. In some scenarios, a cosmological horizon is present, located outside the shell. The energy density and the pressure are given by Eqs. \eqref{sigma0} and \eqref{pres0}, respectively. We distinguish between the configurations where the matter for the conservative perfect fluid at $\Sigma $ satisfies the weak energy condition (WEC), i.e. $\sigma _0 \geq 0$ and $\sigma _0 +p _0 \geq 0$, from those that do not. From Eq. (\ref{potD2VSinSigmaP}) we can determine the stability of the static configurations by using that $V''(a_0)>0$ corresponds to the stable ones.

For the case A, we consider in our analysis the exponent $p=-5$ in Eq. \eqref{metA1}, as a representative example. There exists a critical value of charge $Q_c$ (where the number of horizons in the original outer manifold changes) which plays an important role. For $\Lambda_0>0$, if $Q=0$ the metric has an event horizon, for $0<|Q|<Q_c$ it has the inner and the event horizons, and when $|Q|=Q_c$ they fuse into one to finally disappear if $|Q|>Q_c$, resulting in a naked singularity at the origin; in addition, there is always a cosmological horizon. For $\Lambda_0<0$, with $Q=0$ it has an event horizon, if $0<|Q|<Q_c$ the inner and the event horizons are both present, when $|Q|=Q_c$ they merge, and finally if $|Q|>Q_c$ there is a naked singularity and no horizons. As mentioned above, when $\Lambda _1=0$ we recover the Reissner-Norsdtr\"om with cosmological constant $\Lambda _0$ solution of GR. The electric field equation \eqref{efA1} forces $\Lambda _1<0$.

For the case B, we take $b/M=1$ as a representative value; the horizon structure  is similar to the case A, the differences lie on the values of $Q_c$ and the horizon radii $r_h$ and $r_c$, when they are present. Again, for $\Lambda _1=0$ we recover the Reissner-Norsdtr\"om with cosmological constant $\Lambda _0$  geometry of GR. Now, the electric field equation \eqref{efB} forces $\Lambda _1>0$.

We present the results graphically in Figs. \ref{fig:fig1}, \ref{fig:fig2}, and \ref{fig:fig3}, displaying the most representative of them. In all figures, the top row shows for comparison the results corresponding to GR, i.e. the outer part has the Reissner-Norsdtr\"om with a cosmological constant $\Lambda _0$ geometry. The stability regions for the selected example of the case A metric are displayed in the center row, while those of the case B in the bottom row. All quantities are adimensionalized with the mass. In order to keep our analysis as general as possible, we extend the values of the parameters beyond the range that is physically expected\footnote{Note that large adimensionalized $|\Lambda_1|$ results in a considerable departure from Maxwell electrodynamics.}. We let $\eta$ be outside the interval $0\le \eta <1$; for illustrative purposes the adimensionalized absolute values of $\Lambda _0$ and $\Lambda _1$ used in the plots are quite large. Configurations in the light gray zone are stable, the dashed areas represent thin shells made of matter that does not satisfy WEC, and the dark gray region is non-physical (i.e. $a_0 \le r_h$). In all cases, the most interesting results come when the charge $|Q|$ is close to the critical value $Q_c$. In Fig. \ref{fig:fig1}, where $\Lambda_0 =0$, the three spacetimes are asymptotically flat, in Fig. \ref{fig:fig2}, with $\Lambda_0 <0$ they are asymptotically anti-de Sitter, while in Fig. \ref{fig:fig3}, in which $\Lambda_0 >0$, they are asymptotically de Sitter. In this last figure, the range of $a_0/M$ displayed in the plots does not include the cosmological horizon and the nonphysical region beyond it.

From Figs. \ref{fig:fig1}, \ref{fig:fig2}, and \ref{fig:fig3}, we can see that the value of $Q_c/M$ and the sign of $\Lambda _0$ play a crucial role in GR as well as in both cases of UG. The main features shown in the plots are:
\begin{itemize}
\item In the three spacetimes considered:
\begin{itemize}
\item For any $\Lambda _0 M^2$, as $|Q|/M$ increases, the minimum allowed value of $a_0/M$ shrinks and the stability region grows; there is a dramatic change at $Q_c/M$, from which an arbitrary small value of $a_0/M>0$ is possible and a very large stability region is found. 
\item For $\Lambda _0 = 0$, if $0\le |Q|/M\le Q_c/M$ the matter at the shell always satisfies WEC, but when $|Q|/M >Q_c/M$ it does not for small values of $a_0/M$. For $\Lambda _0 > 0$, if $0\le |Q|/M\le Q_c/M$ the matter at the shell satisfies WEC for small values of $a_0/M$, but not for large ones; when $|Q|/M >Q_c/M$ it satisfies WEC for an intermediate range of values of $a_0/M$, but not for smaller or larger ones. The zone not fulfilling WEC close to the cosmological horizon is not shown in the plots and it corresponds to unstable configurations. For $\Lambda _0 < 0$, if $0\le |Q|/M\le Q_c/M$ the matter only satisfies WEC for small values of $a_0/M$, and if $|Q|/M >Q_c/M$ there is one intermediate zone with the matter fulfilling WEC and two zones in which it does not, one for small values of $a_0/M$ and the other one for large $a_0/M$. 
\item For $\Lambda _0 \ge 0$, if $0\le |Q|/M\le Q_c/M$ stability requires $\eta >0$, but when $|Q|/M >Q_c/M$ any value of $\eta $ is allowed if $a_0/M$ is small enough. For $\Lambda _0 < 0$ stability with any value of $\eta$ is possible, but $\eta <0$ also requires matter not fulfilling WEC if $0\le |Q|/M\le Q_c/M$.
\end{itemize} 
\item A comparison of the three spacetimes shows:
\begin{itemize}
\item The value of $Q_c/M$, where an important change in behavior occurs, is smaller in UG case A than in GR and larger in UG case B than in GR.
\item For fixed values of $\Lambda _0 M^2$ and $|Q|/M$, it seems that in most scenarios  the stability regions are slightly larger in both UG cases than in GR.
\item As mentioned above, both UG cases reduce to GR if $\Lambda_ 1 =0$. In our exploration with values of $\Lambda_ 1$ different from those adopted in Figs. \ref{fig:fig1}, \ref{fig:fig2}, and \ref{fig:fig3}, we have found similar features and a progressive departure from the GR results as $|\Lambda_ 1|$ increases.
\end{itemize}
\end{itemize}
In brief, within UG the non-conservation of the energy-momentum tensor in the outer part of the spacetime, which results in the presence of an extra term (proportional to $\Lambda_1$) in the electric field and in the metric function there, makes that the matter content and the stability regions of the thin shells are different from those corresponding to GR ($\Lambda_1$=0). These differences are small for reasonable values of adimensionalized $\Lambda_1$ and grow with the absolute value of this parameter.

\section{Conclusions}\label{conclu}

We have presented a wide class of spherical spacetimes with thin shells within the theory of UG, constructed by using the junction conditions introduced in \cite{ellis2011}. We have found the matter content for a conservative perfect fluid at the shell and the condition for the stability of the static configurations under perturbations preserving the symmetry. For simplicity, we have taken a linearized equation of state at the shell in our stability analysis. In particular, we have applied this procedure to obtain a charged thin shell with an inner vacuum Minkowski region and an outer zone with a radial electric field. The whole spacetime does not have event horizons or singularities, but in some scenarios it has a cosmological horizon outside the shell. The energy-momentum tensor of the solution adopted for the outer part is non-conservative and we have considered two different possibilities for it, following the cases of spherically symmetric solutions found in \cite{fabris23}. This kind of study is novel in the literature. We have found analytical expressions for all relevant quantities and we have presented their outcome graphically for an easier comprehension. We have compared our results with those corresponding to the GR counterpart, in which the outer zone is the Reissner-Nordstr\"om with a cosmological constant metric. We have found a similar behavior for the matter content and for the stability of the shell in GR and in both UG geometries adopted in our construction. However, some differences arise. We can mention that the value of $Q_c/M$, for which the main change in the behavior takes place, is smaller in UG case A and larger in UG case B compared to GR. Also, for given values of $\Lambda _0 M ^2$ and $|Q|/M$, the stability regions in the plane $(a_0 /M, \eta )$ are slightly larger in most of the UG cases considered here than in the GR counterpart. For a given value of $M$, there is a growing deviation from GR as the value of $|\Lambda_1|$ increases. The new solutions coming from the non-conservation in UG lead to  these differences in the properties of thin shells constructed from them.

\section*{Acknowledgments}

This work has been supported by CONICET. We thank J. C. Fabris for useful discussions on charged black hole solutions in UG.  \\



\end{document}